\documentclass[twocolumn,secnumarabic,amssymb,amsmath,nofootinbib,tightenlines,showpacs,nobibnotes,aps,prl]{revtex4}
\usepackage{amsfonts}
\usepackage{docs}%
\usepackage{bm}%
\usepackage{subfigure}%
\usepackage[colorlinks=true,linkcolor=blue]{hyperref}%
\usepackage{graphicx}
\usepackage{dcolumn}

\begin{document}

\title{Phase Diagram of Two-dimensional Polarized Fermi Gas With Spin-Orbit Coupling}

\author{Xiaosen Yang}
\author{Shaolong Wan}
\altaffiliation{Corresponding author}
\email{slwan@ustc.edu.cn}
\affiliation{Institute for Theoretical Physics and Department of
Modern Physics University of Science and Technology of China,
Hefei, 230026, \textbf{P. R. China}}
\date{\today}

\begin{abstract}
We investigate the ground state of the two-dimensional polarized
Fermi gas with spin-orbit coupling and construct the phase diagram
at zero temperature. We find there exist phase separation when the
binding energy is low. As the binding energy increasing, the
topological nontrivial superfluid phase coexist with topologically
trivial superfluid phase which is topological phase separation.
The spin-orbit coupling interaction enhance the triplet pairing
and destabilize the phase separation against superfluid phase.
\end{abstract}


\maketitle

\section{Introduction}
The topological properties have been investigated extensively in
condensed matter systems such as topological
insulators(TIs)\cite{X.l.Qi, C. L. Kane}, topological
superconductors(TSCs)\cite{N. Read, Zhong Wang, schnyder, Tewari},
etc, which are described by topological order \cite{X. G. Wen}
instead of the traditional Landau symmetry breaking theory. In
ultracold atomic system, the effective spin-orbit coupling(SOC)
has been realized recently by utilizing the spatial varying laser
fields\cite{Y. J. Lin1, Y. J. Lin2}.  With the technique of
Feshbach resonance\cite{Zwierlein, Schunck}, the spin-orbit
coupled ultracold atomic systems provide a clean platform to
investigate the topological properties of the condensed matter
system.

The SOC significantly changes the Fermi surface and largely
enhances the low energy density of state\cite{Vyasanakere,
Salasnich}. Therefor, many interesting phases and intriguing
phenomena become possible. The triplet pairing and the transition
temperature are largely enhanced\cite{Z. Q. Yu} while the pair
coherence lengths are suppressed by the SOC\cite{B B Huang}. In
three dimensions, the ground state of the Fermi system is enriched
by the SOC\cite{Wei yi, Ming gong, Iskin, H. Hu, L. Han, G. Chen,
L. Jiang, Jia Liu}. In two dimensions, the superfluid phase of the
spin-orbit coupled Fermi gas can be topologically
nontrivial\cite{S. L. Zhu, M. Sato, M. sato1, A. Kubasiak}.
Furthermore, there is topologically nontrivial phase
separation(TPS) which is the coexistence of superfluid phases with
different topological order in the trapped SOC Fermi systems with
population imbalance\cite{X yang, J zhou}.

In this paper, we investigate the uniform polarized
two-dimensional(2D) Fermi gas with SOC near a wide Feshbach
resonance at zero temperature. The phase separation is possible
for a polarized Fermi gas without the SOC due to the competition
between the polarization and the pairing interaction. To map out a
exact phase diagram, we determined the ground state by minimizing
the thermodynamic potential of the phase separation\cite{Sheehy}.
In the presence of SOC, the Fermi surface is topologically
changed. The topological phase transition(TPT) takes place when
the excitation gap is closing. Therefor, the topologically
nontrivial superfluid phase(TSF) shows up in the phase diagram
against the topologically trivial superfluid phase(NSF). For the
phase separation phase, the topological phase transition much more
tend to take place in the smaller pairing gap component state,
thus the phase separation becomes topologically nontrivial.

This paper is organized as follows. In Sec.\ref{sec2}, introducing
the Hamiltonian of 2D uniform polarized Fermi gas, we obtain the
zero temperature thermodynamic potential by mean field theory, and
then give the gap equation and the number equations for superfluid
phase. In Sec.\ref{sec3}, we investigate the ground state by
minimizing the thermodynamic potential of the phase separation
phase and map out the phase diagram in detail. A brief conclusion
is given in Sec.\ref{sec4}.

\section{Formalism of the System}
\label{sec2}

We consider the uniform 2D polarized Fermi gas with SOC, which is
described by the Hamiltonian:
\begin{eqnarray}
H=H_{0}+H_{SO}+H_{int}, \label{1}
\end{eqnarray}
where $H_{0}$ is the kinetic term, $H_{SO}$ is the spin-orbit
interaction, and $H_{int}$ is the s-wave interaction between the
two fermionic species. They take
\begin{eqnarray}
&H_{0}&=\sum_{\textbf{k},\sigma} \xi_{\textbf{k},\sigma} c_{\textbf{k},\sigma}^{\dag} c_{\textbf{k},\sigma}, \nonumber\\
&H_{SO}&=\sum_{\textbf{k}} \lambda k \left(e^{-i \varphi_{\textbf{k}}} c_{\textbf{k},\uparrow}^{\dag} c_{\textbf{k},\downarrow} + h.c. \right), \nonumber\\
&H_{int}&= -g
\sum_{\textbf{k},\textbf{k}'}c_{\textbf{k},\uparrow}^{\dag}
c_{-\textbf{k},\downarrow}^{\dag} c_{-\textbf{k}',\downarrow}
c_{\textbf{k}',\uparrow}, \label{2}
\end{eqnarray}
where $\xi_{\textbf{k},\sigma} =\hbar k^{2}/(2m)-\mu_{\sigma}$,
$c_{\textbf{k},\sigma}^{\dag}(c_{\textbf{k},\sigma})$ denotes the
creation(annihilation) operators for a fermion with momentum
$\textbf{k}$ and spin $\sigma=\{ \uparrow, \downarrow \}$,
$\lambda$ is the strength of Rashba spin-orbit coupling,
$\varphi_{\textbf{k}} = \arg(k_{x}+ i k_{y})$, $g$ is the bare
s-wave interaction strength which can be renormalized by
\begin{eqnarray}
\frac{1}{g}=-\sum_{\textbf{k}}\frac{1}{2 \epsilon_{\textbf{k}} +
E_{b}}. \label{3}
\end{eqnarray}

By the transformation,
\begin{eqnarray}
\left(
  \begin{array}{c}
    c_{\textbf{k},\uparrow}\\
    c_{\textbf{k},\downarrow}\\
  \end{array}
\right)=\frac{1}{\sqrt{2}}\left(
  \begin{array}{cc}
    1 & e^{i \varphi_{\textbf{k}}}\\
    e^{-i \varphi_{\textbf{k}}} & -1\\
  \end{array}
\right) \left(
  \begin{array}{c}
    a_{\textbf{k},+}\\
    a_{\textbf{k},-}\\
  \end{array}
\right), \label{4}
\end{eqnarray}
the Eq.\ref{2} becomes
\begin{eqnarray}
&&H_{0}+H_{SO} = \sum_{\textbf{k}, s=\pm} \left(\xi_{\textbf{k},s}
a_{\textbf{k},s}^{\dag} a_{\textbf{k},s}
- h e^{i s \varphi_{\textbf{k}}} a_{\textbf{k},s}^{\dag} a_{\textbf{k},-s}\right), \nonumber\\
&&H_{int} = \sum_{\textbf{k},s=\pm}\left(\frac{\Delta}{2}e^{i s
\varphi_{\textbf{k}}} a_{\textbf{k},s}^{\dag}
a_{-\textbf{k},s}^{\dag} + h.c.\right) + \frac{\mid \Delta
\mid^{2}}{g}, \label{5}
\end{eqnarray}
where $a^{\dag}_{\textbf{k},\pm}(a_{\textbf{k},\pm})$ is the
creation(annihilation) operator for the state with helicity
$(\pm)$, $\xi_{\textbf{k},\pm}=\xi_{\textbf{k}} \pm  \lambda k$
with $\xi_{\textbf{k}}=\epsilon_{\textbf{k}}-\mu$ and the chemical
potentials $\mu=(\mu_{\uparrow}+\mu_{\downarrow})/2$,
$h=(\mu_{\uparrow}-\mu_{\downarrow})/2$, $\Delta$ is the pairing
potential which takes $\Delta= g \sum_{\textbf{k}}
<c_{-\textbf{k},\downarrow} c_{\textbf{k},\uparrow}>$.

The Hamiltonian\ref{1} can be rewritten in the helicity basis
$\Psi_{{\bf k}} = (a_{\textbf{k},+}, a_{\textbf{k},-},
a_{-\textbf{k},+}^{\dag}, a_{-\textbf{k},-}^{\dag})^{T}$ as:
\begin{eqnarray}
H=\frac{1}{2}\sum_{{\bf k}}\Psi_{{\bf k}}^{\dag}\mathcal{H}({\bf k})\Psi_{{\bf k}}
 + \sum_{\textbf{k}}\xi_{\textbf{k}} + \frac{\mid \Delta \mid^{2}}{g}, \label{6}
\end{eqnarray}
with
\begin{eqnarray}
\mathcal{H}({\bf k})=\left(
  \begin{array}{cccc}
    \xi_{\textbf{k},+} & e^{i \varphi_{\textbf{k}}} h  & \Delta e^{i \varphi_{\textbf{k}}} &  0  \\
    e^{-i \varphi_{\textbf{k}}} h & \xi_{\textbf{k},-}  & 0 & \Delta e^{-i \varphi_{\textbf{k}}} \\
    \Delta e^{-i \varphi_{\textbf{k}}} & 0 & -\xi_{\textbf{k},+} & e^{-i \varphi_{\textbf{k}}} h \\
    0 & \Delta e^{i \varphi_{\textbf{k}}} & e^{i \varphi_{\textbf{k}}} h &  -\xi_{\textbf{k},-}  \\
  \end{array}
\right). \label{7}
\end{eqnarray}

We know that the classification of above 2D BdG Hamiltonian, which
breaks the time-reversal symmetry but preserves the particle-hole
symmetry, is $Z$ class \cite{schnyder}. The topological numbers
which characterize the topological properties of the superfluid
phases are integer. There is topological phase transition at the
gap closing point $h= \sqrt{\mu^{2} + \Delta^{2}}$. The
topologically nontrivial superfluid phase show up when $h>
\sqrt{\mu^{2} + \Delta^{2}}$.

The Hamiltonian can be diagonalized as
\begin{eqnarray}
H=\sum_{\textbf{k},s=\pm}
E_{\textbf{k},s}\alpha_{\textbf{k},s}^{\dag}\alpha_{\textbf{k},s}
+ \frac{1}{2} \sum_{\textbf{k},s=\pm}
(\xi_{\textbf{k}}-E_{\textbf{k},s}) + \frac{\mid \Delta
\mid^{2}}{g}, \label{8}
\end{eqnarray}
where, $\alpha_{\textbf{k},\pm}^{\dag}(\alpha_{\textbf{k},\pm})$
is the creation(annihilation) operator for the quasiparticles with
the excitation spectra $E_{\textbf{k},\pm}=
\sqrt{\xi_{\textbf{k}}^{2} + h^{2} + \mid \Delta \mid^{2} +
\lambda^{2} k^{2} \pm 2 E_{0}}$, here $E_{0}= \sqrt{h^{2}
(\xi_{\textbf{k}}^{2}+\mid \Delta \mid^{2}) + \lambda^{2} k^{2}}$.

The thermodynamical potential is $\Omega = -  \text{Tr}
\ln[e^{-\beta H}]$ with $\beta=1/(k_{B} T)$. At $T=0$, the
thermodynamical potential is
\begin{eqnarray}
\Omega =
\frac{1}{2}\sum_{\textbf{k},s=\pm}(\xi_{\textbf{k}}-E_{\textbf{k},s})
+ \frac{\mid \Delta \mid^{2}}{g}. \label{9}
\end{eqnarray}

The pairing gap should be self-consistently determined with
chemical potential by minimizing the thermodynamic potential
$\partial \Omega / \partial \Delta=0$ and the particle number
equations $n_{\sigma}=-\partial \Omega / \partial \mu_{\sigma}$.
They are given as
\begin{eqnarray}
\sum_{\textbf{k}} \frac{1}{2 \epsilon _{\textbf{k}} + E_{b}} =
\frac{1}{4} \sum_{\textbf{k},s=\pm} \frac{1}{E_{\textbf{k},s}}
\left(1+ s \frac{ h^{2}}{E_{0}}\right), \label{10}
\end{eqnarray}
\begin{eqnarray}
n &=& \frac{1}{2}\sum_{\textbf{k},s=\pm} \left[1-\left(1 + s
\frac{h^{2} + \lambda ^{2} k ^{2}}{E_{0}}\right)
\frac{\xi_{\textbf{k},s}}{E_{\textbf{k},s}} \right], \nonumber\\
p n &=& -\frac{1}{2}\sum_{\textbf{k},s=\pm}
\frac{h}{E_{\textbf{k},s}} \left(1 + s \frac{
\xi_{\textbf{k},s}^{2} +\Delta ^{2} }{E_{0}} \right) , \label{11}
\end{eqnarray}
where, $n=n_{\uparrow}+n_{\downarrow}$ is the total particle
number and $p=(n_{\uparrow}-n_{\downarrow})/n$ is the
polarization. In the presence of SOC, the Fermi surface is
topologically changed and the triplet pairing is possible. The
condensate fraction should include singlet and triplet
contributions $n_c = n_{0}+n_{1}$ which are given as
\begin{eqnarray}
n_{0} &=& 2 \sum_{\textbf{k}} \mid < c_{\textbf{k},\uparrow} c_{-\textbf{k},\downarrow} > \mid ^{2} \nonumber\\
&=&\frac{\Delta ^{2}}{8}\sum_{\textbf{k}} \left[ \sum_{s=\pm}
\left(1 + s \frac{h^{2}}{E_{0}} \right)
\frac{1}{E_{\textbf{k},s}}\right]^{2}, \label{12}
\end{eqnarray}
\begin{eqnarray}
n_{1} &=& \sum_{\textbf{k}} (\mid <c_{\textbf{k},\uparrow} c_{-\textbf{k},\uparrow} > \mid ^{2}
+ \mid <c_{\textbf{k},\downarrow} c_{-\textbf{k},\downarrow} > \mid ^{2}) \nonumber\\
&=&\frac{\Delta ^{2}}{16} \sum_{\textbf{k}}\left[
\left(\sum_{s=\pm} \frac{s 1}{E_{\textbf{k},s}} \right)^{2}
\sum_{s=\pm} \frac{\lambda^{2} k^{2}(\xi_{\textbf{k}}+s
h)^{2}}{E_{0}^{2}} \right]. \label{13}
\end{eqnarray}

\section{The Phase Diagram in $p-\lambda k_F /E_F$ Plane}
\label{sec3}

There is no guarantee that the ground state of the polarized Fermi
gas corresponds to one of the spatially homogeneous states. As the
competition between the population imbalance and the pairing interaction,
the phase separation becomes possible. For the polarized Fermi gas, the stability of the phase separation
against the superfluid should be considered like the case without
SOC. By introducing the mixing coefficient $x (0\leq x \leq1)$ and
ignoring the interfaces energy between the two coexisting phase,
the thermodynamic potential of the phase separation can be written
as
\begin{eqnarray}
\Omega  = x \Omega(\Delta_{1}) + (1-x) \Omega(\Delta_{2}),
\label{14}
\end{eqnarray}
where, $\Delta_{i} (i=1,2)$ is the pairing gap of the $i$
component separated state. The thermodynamic potential should be
minimized with $\Delta_{i}$ and the mixing coefficient $x$. The
number equations become $n_{\sigma} = x n_{\sigma} (\Delta_{1}) +
(1-x) n_{\sigma} (\Delta_{2})$. By solving the gap equations and
the number equations selfconsistently, we construct the phase
diagram in $p-\lambda k_{F}/E_{F}$ plane for different binding
energy.
\begin{figure}
\subfigure{\includegraphics[width=4.0cm, height=3.5cm]{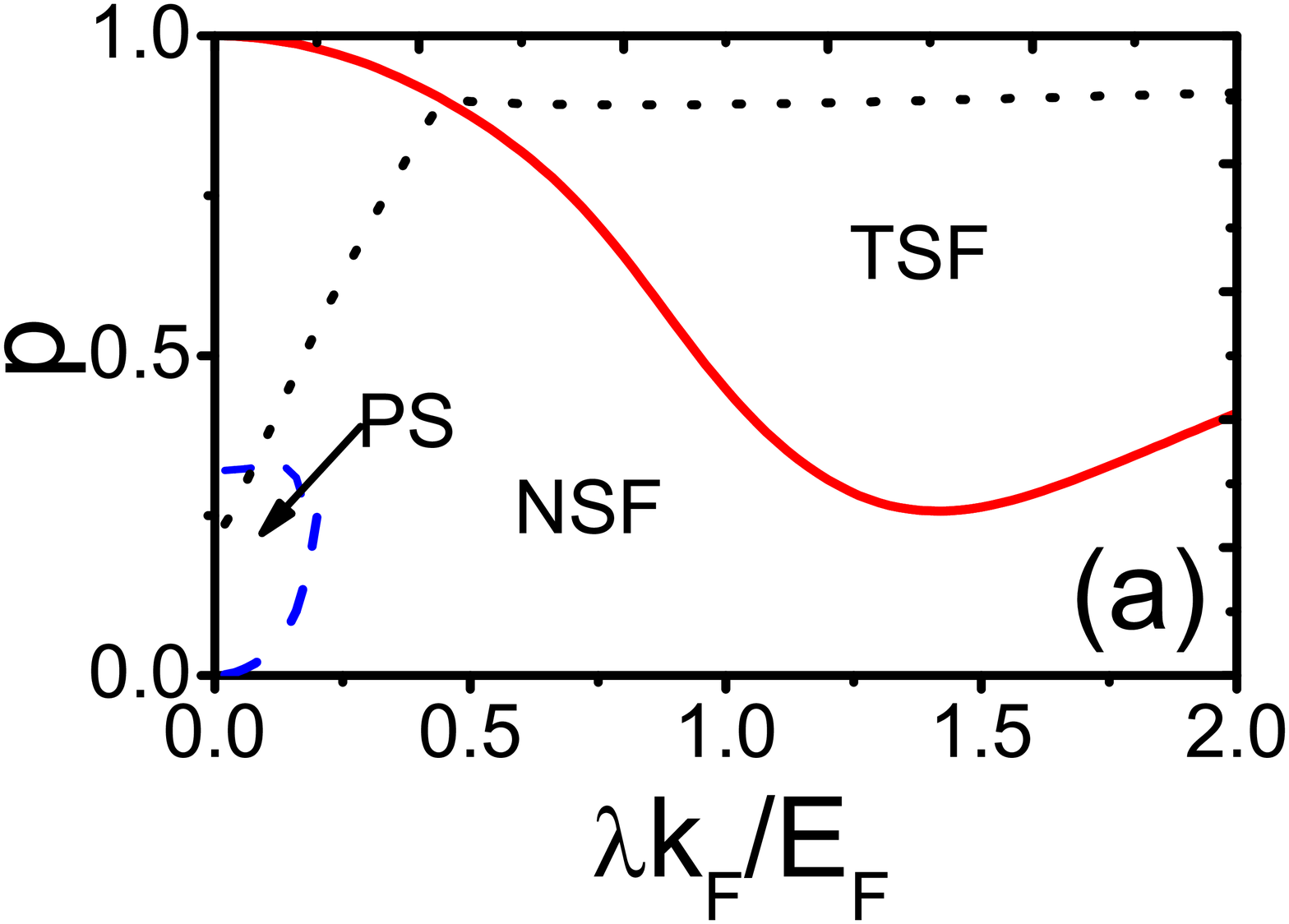}}
\subfigure{\includegraphics[width=4.0cm, height=3.5cm]{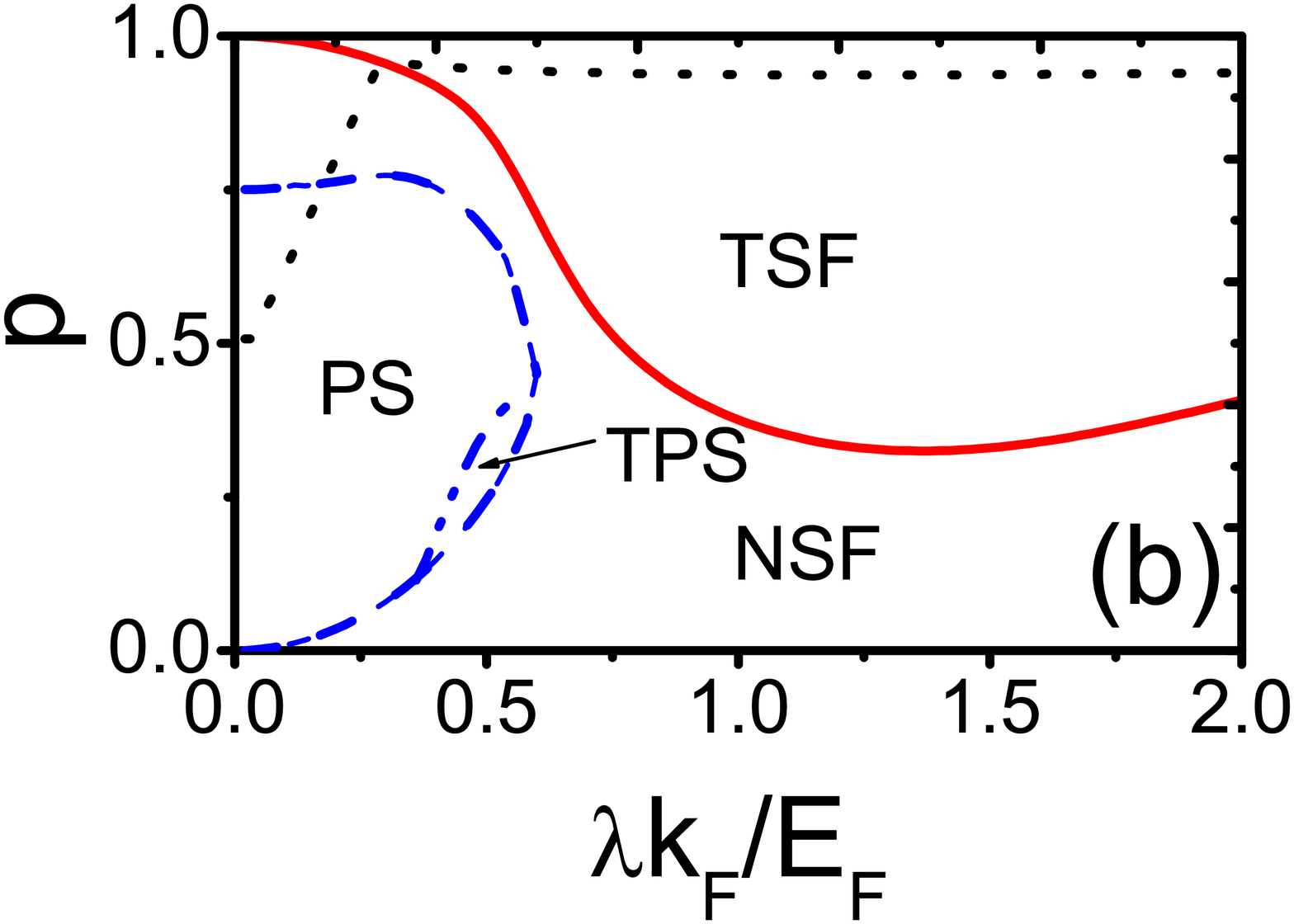}}
\subfigure{\includegraphics[width=4.0cm, height=3.5cm]{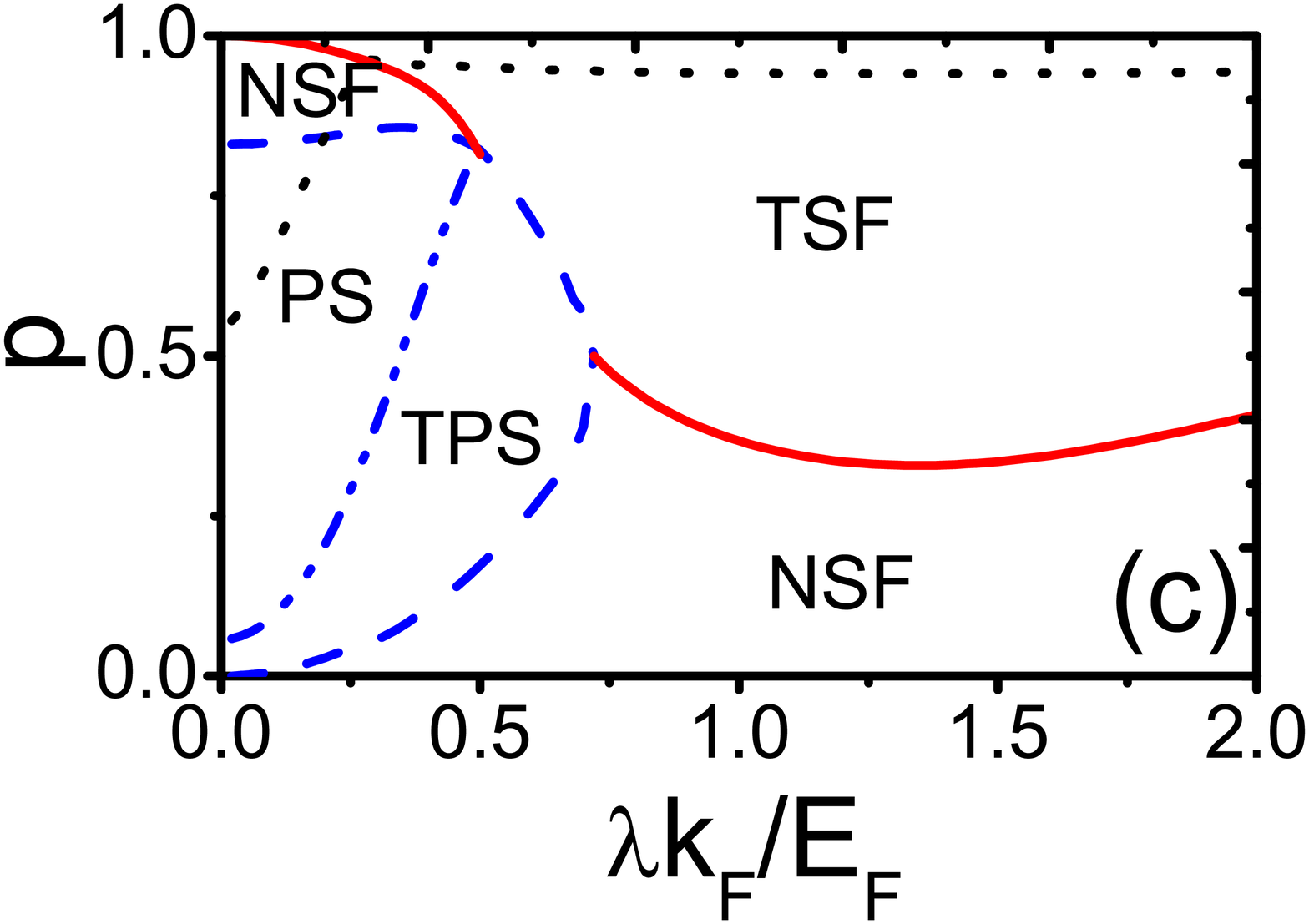}}
\subfigure{\includegraphics[width=4.0cm, height=3.5cm]{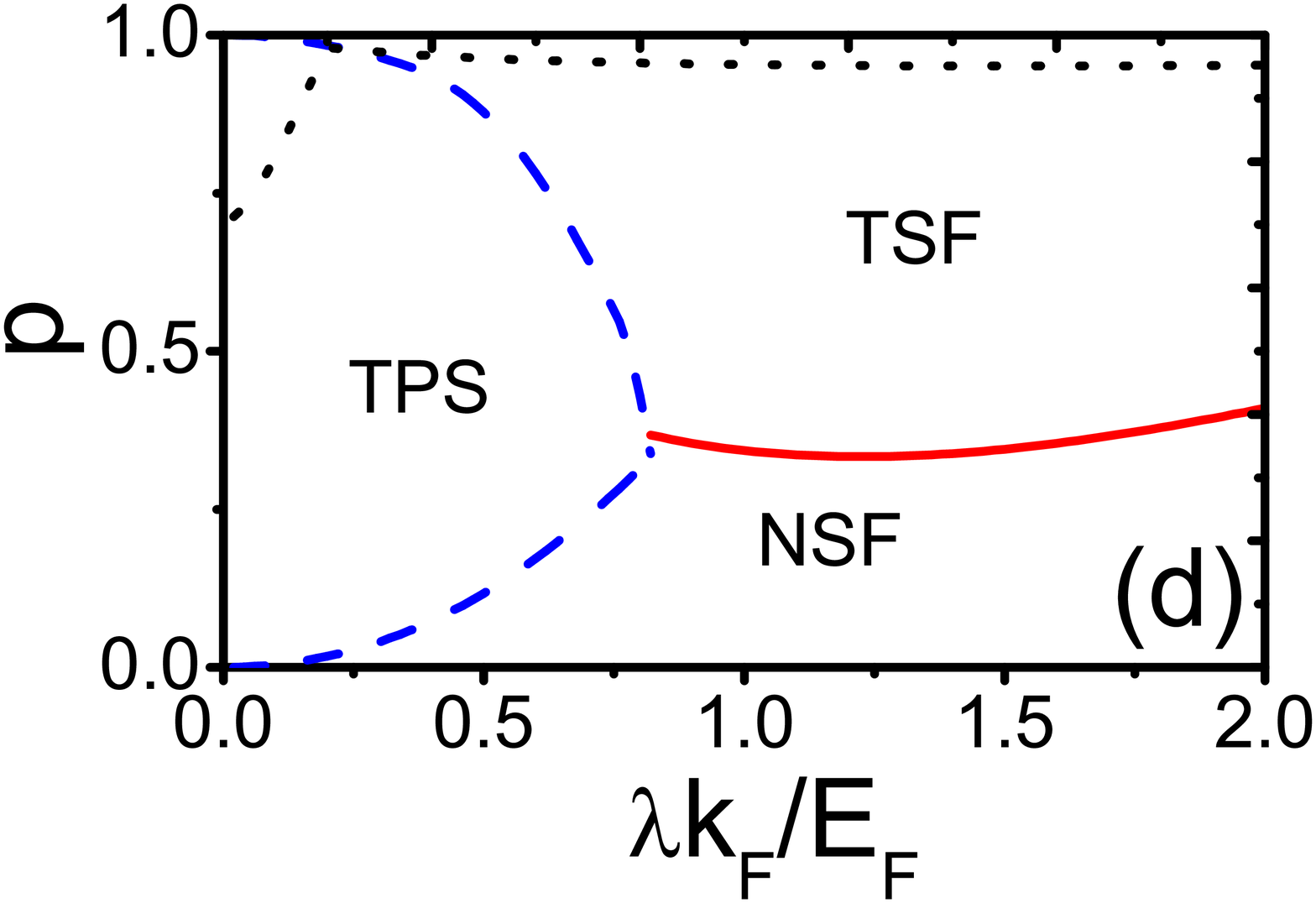}}
\caption{The phase diagrams in $p-\lambda k_F /E_F$ plane with
binding energy (a) $E_b=0.1 E_F$; (b) $E_b=0.5 E_F$; (c) $E_b= 0.6
E_F$; (d) $E_b=1.0 E_F$. Here, $E_{F}= k_{F}^{2}/2m = n \pi /m $.
The red solid lines separate the TSFs from NSFs phases. The blue
dash lines separate the phase separation from superfluid phase.
The blue dash-dot-dot lines are the boundaries between
topologically trivial and nontrivial phase separation. The dot
lines denote the $\Delta / E_F = 10^{-3}$, above which the pairing
gap is lower than $10^{-3}$. } \label{fig.1}
\end{figure}

\begin{figure}
\includegraphics[width=8.0cm, height=6.5cm]{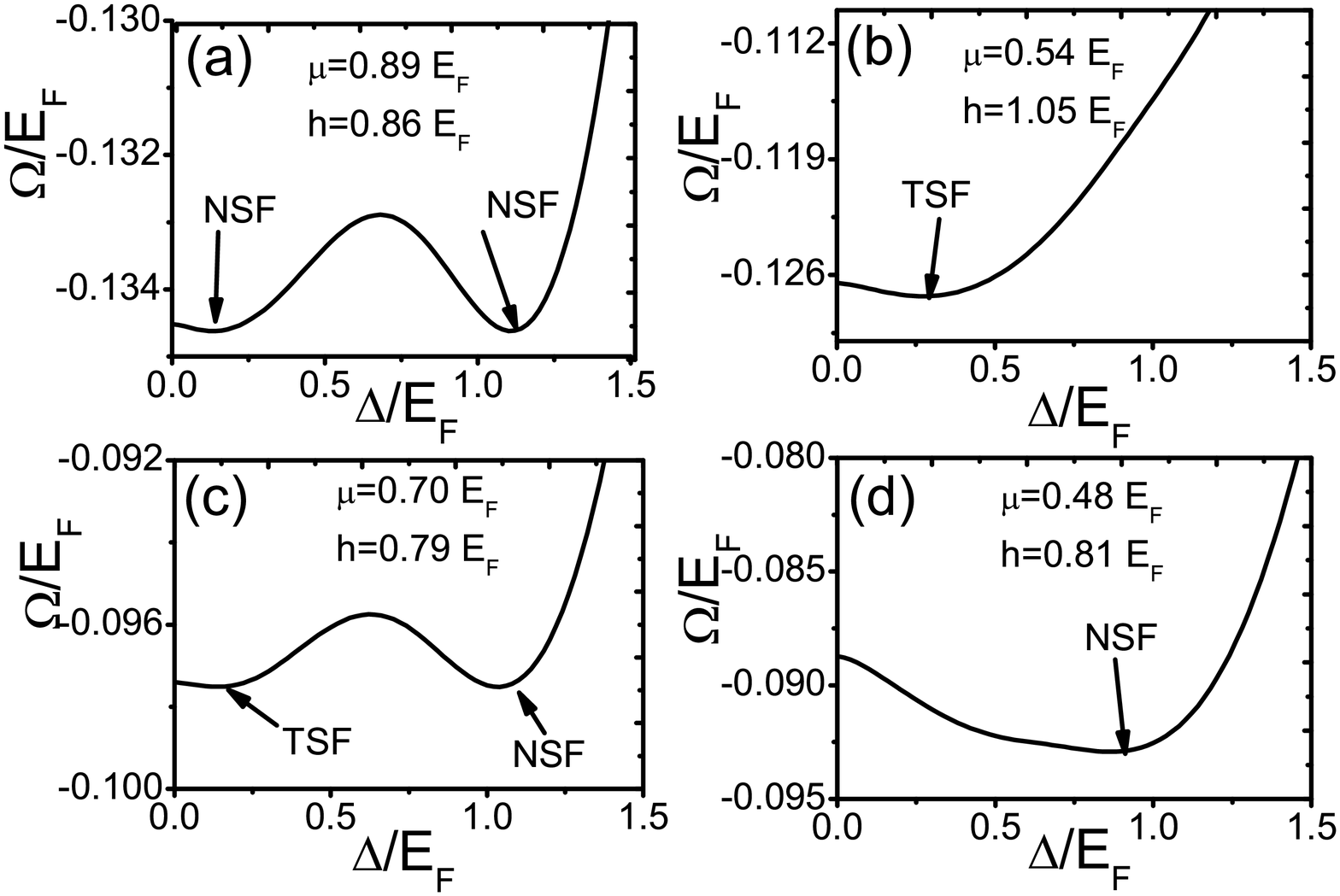}
\caption{The thermodynamic potential $\Omega$ as a function of the
pairing gap $\Delta$ with the binding energy $E_b / E_F = 0.6$ for
(a) $\lambda k_F /E_F = 0.4$, $p=0.8$; (b) $\lambda k_F /E_F =
0.7$, $p=0.8$; (c) $\lambda k_F /E_F = 0.4$, $p=0.3$; (d) $\lambda
k_F /E_F = 0.7$, $p=0.3$.} \label{fig.2}
\end{figure}

\begin{figure}
\subfigure{\includegraphics[width=8.5cm, height=5.0cm]{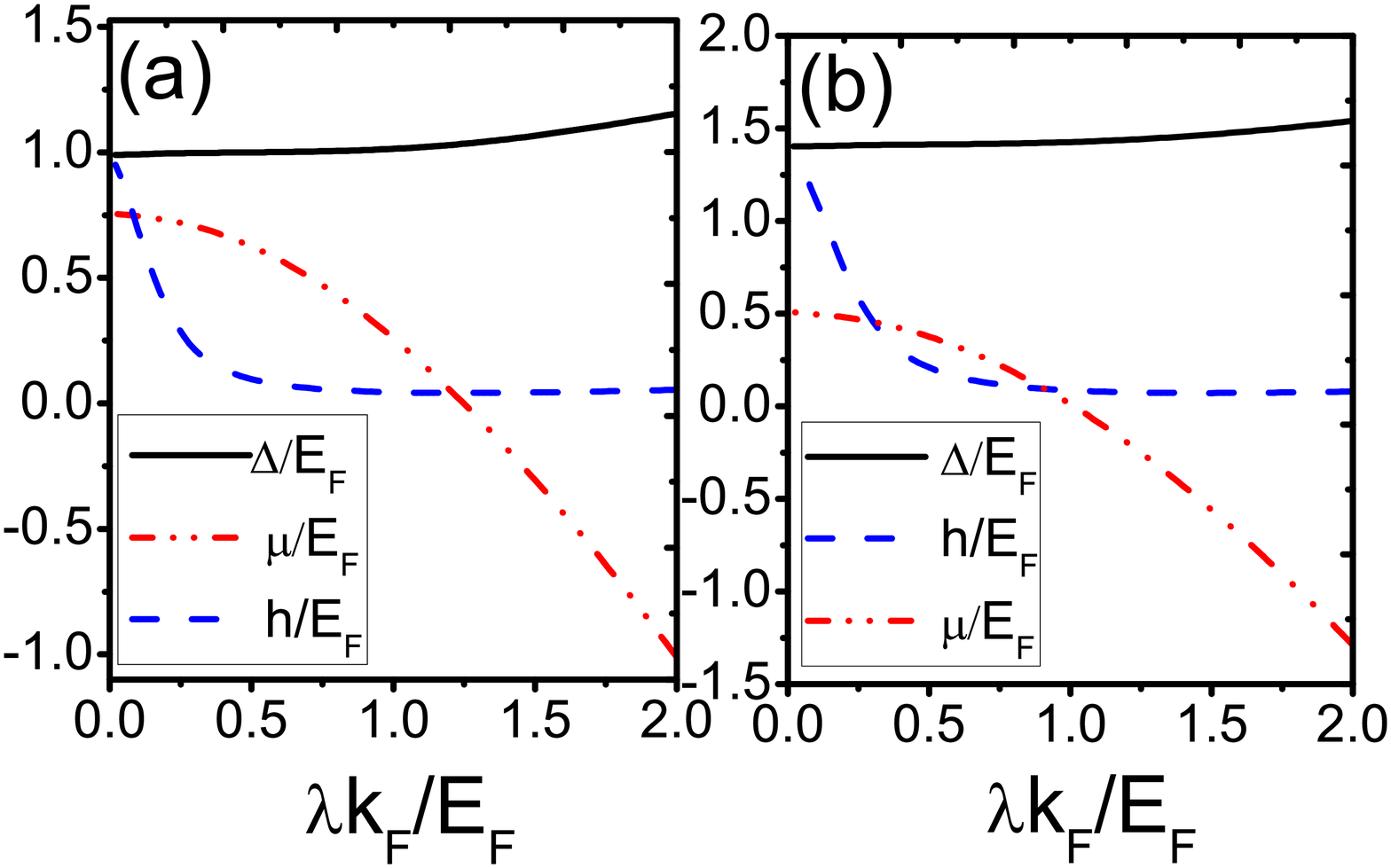}}
\subfigure{\includegraphics[width=8.0cm, height=5.0cm]{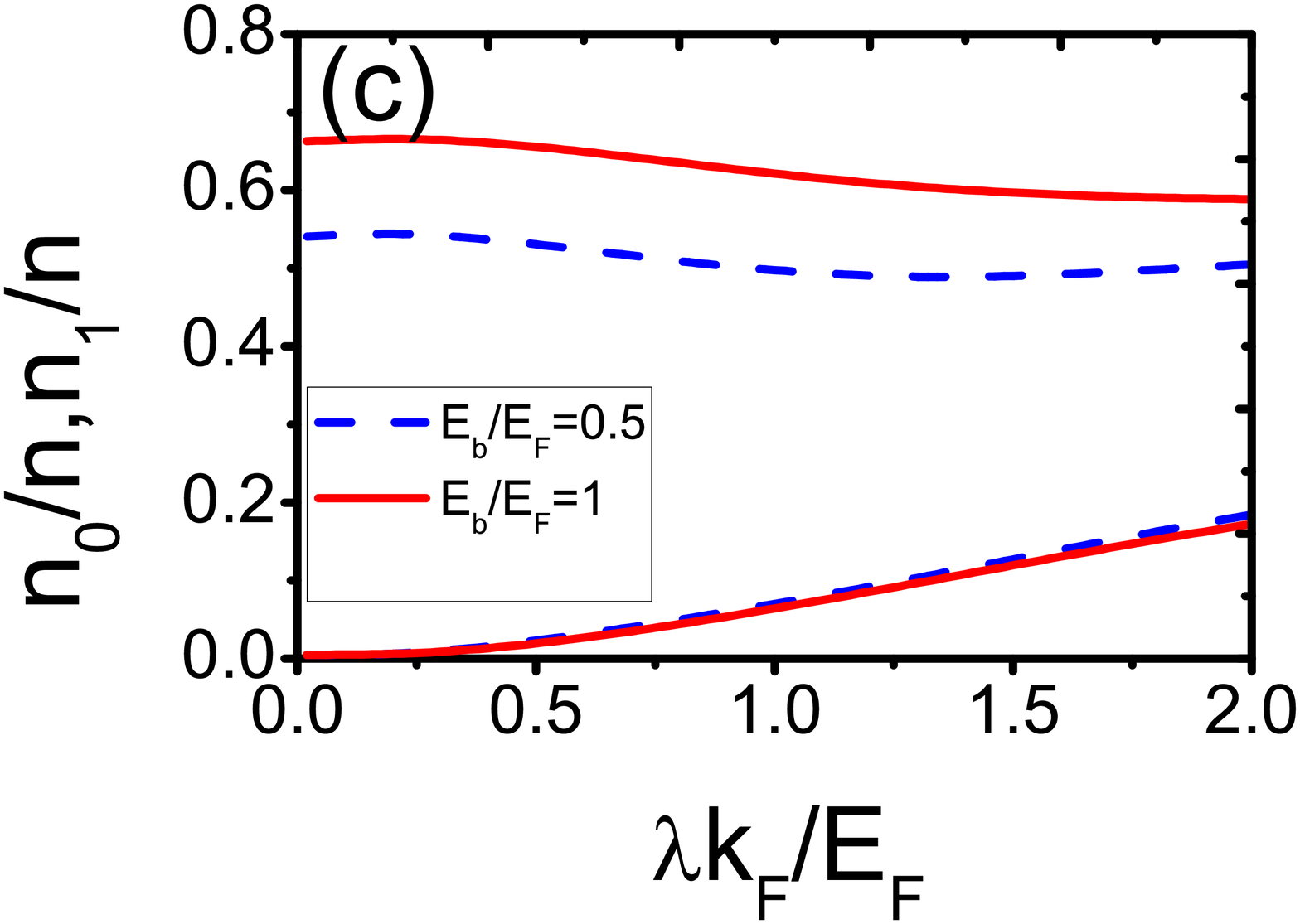}}
\caption{The pairing gap and the chemical potential as functions
of the SOC strength $\lambda k_F /E_F$ with $p=0.01$ for (a)
$E_b=0.5 E_F$ and (b) $E_b=1.0 E_F$. (c) The condensate fractions
of singlet and triplet contribution as functions of the SOC  for
$E_B / E_F = 0.5, 1$ with $p=0.01$. The above two lines are
singlet contributions while the others are triplet contribution.
The triplet contributions are enhanced by the SOC.} \label{fig.3}
\end{figure}

First, we give the phase diagrams in $p-\lambda k_F /E_F$ plane
with different binding energy in Fig.\ref{fig.1}((a) $E_b=0.1
E_F$; (b) $E_b=0.5 E_F$; (c) $E_b= 0.6 E_F$; (d) $E_b=1.0 E_F$).
The phase separation, which is coexistence of distinct
topologically trivial superfluid phases, show up in the absence of
SOC. When the polarization is larger than $0.32$, the phase
separation can not sustain against topologically trivial
superfluid (NSF) in the phase diagram without SOC for $E_b=0.1
E_F$ case. The critical polarization increase with the binding
energy as shown in Fig.\ref{fig.1}. This consist with the recent
result without the SOC \cite{caldas}.

In the presence of SOC, the Fermi surface is topologically changed
and other interesting topologically nontrivial phases are
possible. There is topological phase transition when the
excitation gap closing at the critical point $h= \sqrt{\mu^{2} +
\Delta^{2}}$. The topological phase transition tend to take place
in the high polarization area in which the pairing gap is low and
the imbalance of the chemical potential is large. Therefor, the
phases are TSF in the phase diagrams with high polarization as
shown in Fig.\ref{fig.1}.

For the phase separation phase, the topological phase transition
much more tend to take place in the low pairing gap component
state. The phase separation become topologically nontrivial when
the low pairing gap component state become topologically
nontrivial as shown in Fig.\ref{fig.1}(b),(c),(d). As the binding
energy increasing, the topological phase separation is more
possible. The entire phase separation is topologically trivial
with $E_b=0.1 E_F$(shown in Fig.\ref{fig.1}(a)) and nontrivial
with $E_b=1.0 E_F$(shown in Fig.\ref{fig.1}(d)). The boundary (the
red sold line) between the TSF and NSF merge with the phase
separation boundary (the blue dash line) as the binding energy
increasing. Fig.\ref{fig.1} also show that the SOC destabilize the
phase separation against superfluid phase. When the SOC strength
increase to a critical value, the phase separation disappear.

Second, we show the behavior of thermodynamic potential toward the
pairing gap in different phase regions of the phase diagram for
$E_b / E_F =0.6$ in Fig.\ref{fig.2}. The thermodynamic potential
has two degenerate minimums in the phase separation regions as
shown in Fig.\ref{fig.2}(a)(c). The two distinct superfluid phases
can show up and coexist in the phase diagram. The two coexistent
states are all topologically trivial in Fig.\ref{fig.2}(a). But,
the the smaller component state is topologically nontrivial while
the other is topologically trivial in  Fig.\ref{fig.2}(c). The
thermodynamic potential in the superfluid region only has one
minimum as shown in Fig.\ref{fig.2}(b)(d).

Finally, we show the variation of $\Delta$, $\mu$, $h$ and the
condensate fractions for very low polarization ($p=0.01$) with
$E_b=0.5 E_F$ and $E_b=1 E_F$ in Fig.\ref{fig.3}. The triplet
condensate fractions are enhanced by the SOC. The SOC enhance the
triplet pairing in virtue of the topologically change the Fermi
surface. Therefor, the system can not sustain the phase separation
against the superfluid phase as the triplet pairing increasing as
well as the SOC strength.

It should be point out that the gap equation divergent as the
pairing gap $\Delta$ reduce to zero when $\mu < -(\lambda^{4} + 4
h^{2})/(4 \lambda^{2}) $ or $\mu < \text{min}(-\mid h \mid/2 ,
-\lambda^{2}/2  )$, hence there is no boundary between the normal
phase and the superfluid phase. We map out the boundaries (dot
lines) for $\Delta = 0.001 E_F$ as shown in Fig.\ref{fig.1}. Above
the curve, the pairing gap is $\Delta < 0.001 E_F$ and
exponentially decreases as the SOC reduce to zero\cite{J zhou}.

\section{Conclusions}
\label{sec4}

We construct the phase diagram for the two-dimensional Fermi gas
with spin-orbit coupling and population imbalance near a wide
Feshbach resonance. We map out the stability regions of the
topologically trivial and nontrivial superfluid phase, and phase
separation in detail. As the spin-orbit coupling increasing, there
is topological phase transition. Therefor, the topologically
nontrivial phase separation is possible. The spin-orbit coupling
enhance the triplet pairing and suppress the phase separation. The
phase separation can not sustain against superfluid phase when the
spin-orbit coupling is large.

\begin{acknowledgments}
We are very grateful to Wei Yi and Zhong Wang for helpful
discussions. This work is supported by NSFC Grant No.10675108.
\end{acknowledgments}

\end{document}